\documentclass[12pt,twoside,onecolumn,draftcls]{IEEEtran} 
\usepackage{graphicx,cite,amsmath,amssymb,hhline,booktabs,stfloats,bm}
\usepackage{verbatim}
\usepackage[utf8]{inputenc}
\usepackage{subfigure}
\usepackage{algorithm,algorithmic}

\begin{document}
%
\title{Joint Energy Minimization and Resource Allocation in C-RAN with Mobile Cloud}
%
%
%
%

\author{Kezhi~Wang,~
        Kun~Yang,~\IEEEmembership{Senior Member,~IEEE,}
        and~Chathura~Sarathchandra~Magurawalage
\IEEEcompsocitemizethanks{\IEEEcompsocthanksitem Kezhi Wang, Kun Yang and Chathura Sarathchandra Magurawalage are with the School of Computer Sciences and Electrical Engineering, University of Essex, CO3 4HG,
Colchester, U.K.\protect\\
E-mails: \{kezhi.wang, kunyang, csarata\}@essex.ac.uk.
}
\thanks{Manuscript received July 30, 2015; revised November 25, 2015.}}

%
%

\markboth{IEEE journal,~Vol.~XX, No.~X, XX~XXXX}%
{Shell \MakeLowercase{\textit{et al.}}: Bare Demo of IEEEtran.cls for Computer Society Journals}
%



\IEEEtitleabstractindextext{%
\begin{abstract}
Cloud radio access network (C-RAN) has emerged as a potential candidate of the next generation
access network technology to address the increasing mobile traffic, while mobile cloud computing (MCC) offers a prospective solution to the resource-limited mobile user in executing computation intensive tasks. Taking full advantages of above two cloud-based techniques, C-RAN with MCC are presented in this paper to enhance both performance and energy efficiencies. In particular, this paper studies the joint energy minimization and resource allocation in C-RAN with MCC under the time constraints of the given tasks. We first review the energy and time model of the computation and communication. Then, we formulate the joint energy minimization into a non-convex optimization with the constraints of task executing time, transmitting power, computation capacity and fronthaul data rates. This non-convex optimization is then reformulated into an equivalent convex problem based on weighted minimum mean square error (WMMSE).
The iterative algorithm is finally given to deal with the joint resource allocation in C-RAN with mobile cloud. Simulation results confirm that the proposed energy minimization and resource allocation solution can improve the system performance and save energy.
\end{abstract}

\begin{IEEEkeywords}
C-RAN, Joint Energy Minimization, Mobile Cloud Computing, Resource Allocation.
\end{IEEEkeywords}}

\maketitle

\IEEEdisplaynontitleabstractindextext

%
\IEEEpeerreviewmaketitle

\newpage
\section{Introduction}
Nowadays, the number of the smart devices and the
corresponding mobile traffic have grown rapidly, which poses
an increasingly high burden on the existing cellular network. It
is predicted that the mobile device traffic will increase one thousand times and the cost is expected to decrease one hundred times by
2020, with the help of new network and computation paradigm \cite{6824752}. Moreover, more and more computational resource intensive tasks, such as multimedia applications, high definition video playing and gaming appear in our daily life, make the load of both the mobile phone and the network, in terms of energy and bandwidth, increase hugely. Further, those types of applications have the trend of attracting more and more attention from the smartphone users.

However, in traditional cellular networks, each base station (BS) transmits data signal separately to the user equipment (UE), so that the energy cost in the BS will be usually very high, in order to overcome the path loss and the interference from the other BSs. Cooperative relaying has been proposed to mitigate and combat the deleterious effects of fading by sending and receiving independent copies of the same signal at different nodes. However, the total energy cost of the cooperative relaying still may be a little bit high \cite{6898856,6942282}.
Also, coordinated Multi-Point (CoMP) technique has been proposed to mitigate interference by using cooperation techniques, such as joint transmission (JT) and coordinated beamforming (CBF), between different BSs. CoMP technique sometimes cannot achieve the best performance, due to traditional X2 interface limitation, i.e., low bandwidth, high latency and inaccurate synchronization.

It is very fortunate that recently, a new promising network infrastructure, i.e., cloud radio access network (C-RAN), has been presented and soon received a large amount of attention in both academia and industry \cite{6998041,6272415}. C-RAN is a cloud computing based, centralized, clean and collaborative radio access network \cite{China}. It divides the traditional BS into three parts, namely, serval remote radio heads (RRHs), the baseband unit (BBU) pool, and the high-bandwidth, high-speed, low latency fiber transport (or fronthaul) link connecting RRH to the BBU cloud pool. In C-RAN, most of the intensive network computational tasks, such as baseband signal processing, precoding matrix calculation, channel state information estimation are moved to BBU pool in the cloud, which is composed of numerous software defined virtual machines with the feature of dynamically configurable, scalable, sharable, re-allocatable per demand. On the other hand, RRHs, which act as the soft relay, can compress and forward the received signals from the BBU cloud and transmit them in the RF frequency band to UEs.
In this case, RRHs,
with limited functions, only including A/D, D/A conversion, amplification, frequency conversion, make them very easy to distribute, according to the network requirement. Thanks to the separation of BBU and RRH and the cooperation between different BBUs, significant performance gain can be achieved in terms of efficient interference cancellation and management as well as the increase of network capacity and decrease of the energy cost.
The benefits of C-RAN were also given in \cite{6272415} from the industry perspectives.

Another very impressive technique, i.e. mobile cloud computing (MCC) has attracted a huge number of interest recently \cite{6195845,5445167}. MCC is inspired by integrating the popular cloud computing into mobile environment, which enables that mobile user with increasing computing demands but limited computing resource can offload tasks to the powerful platforms in the cloud. The reference \cite{5445167} has investigated if the offloading operation to the cloud can save energy and extend battery lifetimes for UEs. The reference \cite{6574874} has provided a theoretical framework of energy optimal mobile cloud computing under stochastic wireless channel while the reference \cite{6787113} has proposed a game theoretical approach for achieving efficient computation offloading for MCC.
Also, energy-efficiency oriented traffic offloading in wireless networks has been studied in \cite{7012044}. The integration of cloud computing into vehicular networks has been investigated in \cite{6616115}, in which the vehicles can share computation resources, storage
resources and bandwidth resources each other.
Cooperative resource management in cloud-enabled vehicular networks has also been studied in \cite{7275140}, where the resource management between bandwidth and computing resources in cloud-enabled vehicular networks has been considered.
Reference \cite{6863135} has proposed
a cloud-based wireless multimedia social network, where the desktop users can receive multimedia services
from a multimedia cloud, and they also can share their live contents with mobile friends through wireless connections.
Moreover, software defined network (SDN) has been proposed to offer
scalable and flexible management with a logical centralized
control model to MCC \cite{7166189,7321982}.

Although the cloud computing has demonstrated the potential ability to improve the performance, in not only the MCC, but also C-RAN, the research of integration between them is rarely less. Fortunately, \cite{Chathura ,6849260,7105959} have shown that the combination of MCC and C-RAN is of huge interest. Reference \cite{Chathura} has shown that computing resources
and communication resources can be coupled for enhancing connected devices. Reference \cite{6849260} has studied the topology configuration and rate allocation in C-RAN with the objective of optimizing the end-to-end TCP throughput performance of MCC. Reference \cite{7105959} has investigated a cross-layer resource allocation model for C-RAN to minimize the overall system power consumption in both the BBUs and RRHs.

Moreover, pursuing computational intensive or high bandwidth tasks in the UE side increases the operating expense and capital expenditure of the mobile operators, which drastically reduce their profit and make them face a very hard situation. It has been shown that the energy overhead or the electricity cost are among the most important factors in the overall operational expenditure \cite{6133172}. Thus, how to save the whole system's energy is of huge importance and interest in the operators' eyes.

To address the above-mentioned questions, in this paper, we propose a novel C-RAN structure with the mobile cloud (virtual machine) co-located with the BBU in the cloud pool. The mobile cloud is responsible for the execution of the computational intensive task while the BBU is in charge of returning the execution results to the UE via RRHs. We aim to jointly reduce the total energy cost under the time constraints of the given task in C-RAN and mobile cloud.
In particular, we model the energy cost of the mobile cloud in executing the task, and the energy cost of the network in transmitting the results back to UE through RRHs. We also model the time spent in the mobile cloud and in wireless transmission process. We formulate the joint energy minimization into a non-convex optimization, which is NP-hard. Then we convert it to the power minimization plus the sum data rate (throughput) maximization problems.
Sum data rate (throughput) maximization problem can be transformed to
the equivalent minimization of the weighted mean square error (MSE) problem, which can be solved by weighted minimum mean square error (WMMSE) solution \cite{5756489,4712693}.
By using the WMMSE-based iterative algorithm, we can successfully address the joint resource allocation between the mobile cloud and C-RAN and also deal with beamforming vector design in RRHs.

The remainder of this paper is organized as follows. Section
2 introduces the system model including the mobile cloud computational model and the network model.
Section 3 presents the optimization problem formulation as well as two separate energy minimization solutions in mobile cloud and C-RAN, while Section
4 introduces the joint energy minimization algorithm in mobile cloud and mobile network. Simulation results are shown in Section
5, followed by concluding
remarks in Section 6.

\section{System Model}
In this section, the mathematical models for the mobile cloud computation as well as the C-RAN are presented. First, we introduce the concept of the mobile clone in MCC and the whole system design, and then we describe the computation models, including the energy and time consumption model in the cloud and in the network. Finally, the qualify of service (QoS) requirement is given through the time constraint of the given task.

\subsection{Mobile Clone and System Architecture}
Normally, when the mobile users come across the computational intensive or high energy required tasks, they sometimes do not want to offload those tasks into the mobile cloud, as transmitting those program data to the cloud still costs some energy \cite{5445167}. In some cases, it is even better to execute those tasks locally if transmission overhead is too high. Therefore, it is better to have the mobile user's computational tasks and some of the corresponding data in the mobile cloud first. To deal with this concerns, we propose to have \textit{mobile clones} which are co-located with the BBU in the cloud pool. The mobile clone will have the user task information and data on board.
Mobile clone can be implemented by the cloud-based virtual machine which holds the same software stack, such as operating system, middleware, applications, as the mobile user.
If the mobile user wants to execute some task, it only needs to send the indication signal and the corresponding user configuration information to the mobile clone (virtual machine), which will execute those tasks on mobile user's behalf. In this case, the mobile user only needs to cost a small amount of energy and time overhead. After the task execution completion, the mobile clone will transmit the computation result data back to the mobile user through C-RAN. Another advantage of having mobile clone is that each mobile clone can talk to each other in the cloud without through the wireless link. In this case, each mobile user's communication can be possibly transferred into the communication between the mobile clones (clone-to-clone communication), thereby saving a great number of the wireless network resources as well as the energy and time overhead.

\begin{figure*}[htbp]
\centering
\includegraphics[width=5in]{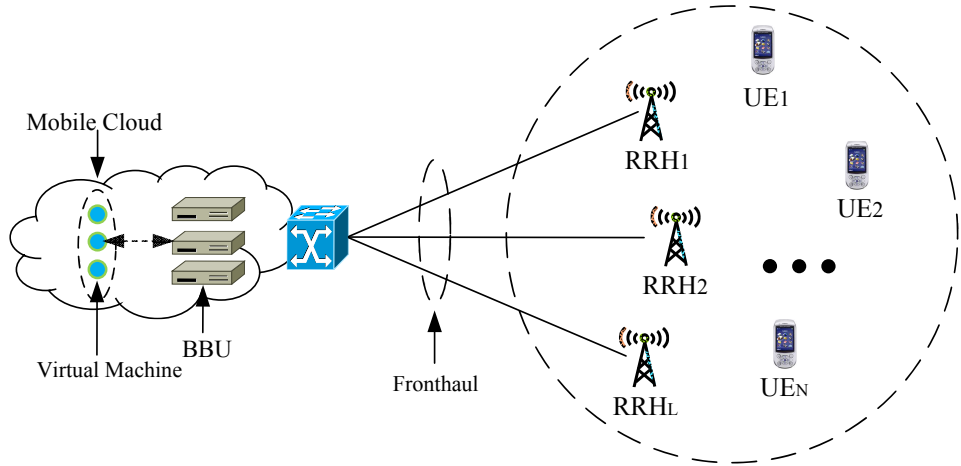}
\caption{A cloud radio access network with mobile cloud system.} \label{figf1}
\end{figure*}

In this paper, we consider there are $\mathcal{N}=\{1, 2,..., N\}$ UEs, each with one antenna, deployed in the C-RAN. Also, we consider there are $\mathcal{L}=\{1, 2,..., L\}$ RRHs, each of which has $K\geq 1$ antennas, connecting to the BBU pool through high-speed fiber fronthaul link, as shown in Fig. \ref{figf1}.
We consider the case that each mobile user already has one specific mobile clone, established in the cloud, beside the BBU, and the mobile clone has the same software stack as its corresponding mobile user.
Similar to \cite{5445167} and \cite{6787113}, we assume that each of UE $i$ has the computational intensive task $U_i$ to be accomplished in the mobile clone $i$ as follows
\begin{equation}\label{en1}
\begin{aligned}
U_i=(F_i, D_i), \;  i=1, 2,..., N
\end{aligned}
\end{equation}
where $F_i$ describes the total number of the CPU cycles needed to be completed for this computational task $U_i$ for the $i$-th UE,
while $D_i$ denotes the whole size of the task's output data transmitting to the $i$-th UE through C-RAN after task execration, including the task's output parameter and the calculation results, etc. $D_i$ and $F_i$ can be obtained by using the approaches provided in \cite{6253581}.

Since the mobile clone has the same software stack as the UE, UE only needs to transmit a small amount of the data including the indication signal and configuration information to the mobile clone to instruct the task to be executed. Therefore, we do not consider the time and energy consumption caused in the uplink transmission process. Also we assume that all the channel state information (CSI) are available in the BBU pool, which facilitate interference cancelation and signal cooperation. We do not consider the time and energy consumption in the fronthaul link, but we will consider the the fonthaul constraints by using the transmitting data rate.

\subsection{Computation Model}
In the mobile clone, the time spent to complete the task $U_i$ is defined as follows
\begin{equation}\label{en2}
\begin{aligned}
T_i^C=\frac{F_i}{f_i^C},  \;\;i=1, 2,...,N
\end{aligned}
\end{equation}
and the energy used in the $i$-th mobile clone is given as
\begin{equation}\label{en3}
\begin{aligned}
E_i^C=\kappa_i^C (f_i^C)^{\nu^C_i-1} F_i,  \;\;i=1, 2,...,N
\end{aligned}
\end{equation}
where $\kappa_i^C \geq 0$ is the effective switched capacitance, $f_i^C$ is the computation capability of the $i$-th virtual machine serving UE $i$ in the cloud and $\nu^C_i \geq 1$ is the positive constant \cite{6748976}. According to the realistic measurements, $\kappa_i^C$ can be set to $\kappa_i^C=10^{-11}$ \cite{Antti}.

We also assume that different mobile clone may have different computational capacity and the constraint of the computation capacity $f_i^C$ for the virtual machine is given by
\begin{equation}\label{en4}
\begin{aligned}
f_i^C \leq  f^C_{i,max},  \;\;i=1, 2,...,N
\end{aligned}
\end{equation}
where $f^C_{i,max}$ is the maximum computation capacity that the $i$-th virtual machine can achieve, as in the reality, the virtual machine normally cannot have unlimited computational capability.

\subsection{Network Model}
After the mobile clone completes the execution of the task, the results will be returned to the mobile user through C-RAN.
The received signal at the UE $i$ under the complex baseband equivalent channel can be written as
\begin{equation}\label{en5}
\begin{aligned}
&y_i=\sum_{j\in \mathcal{C}}\mathbf{h_{ij}}^H \mathbf{v_{ij}} x_i +\sum_{k\neq i}^N \sum_{j\in \mathcal{C}}\mathbf{h_{ij}}^H \mathbf{v_{kj}} x_k+ \sigma_i, \\&  i=1, 2,...,N
\end{aligned}
\end{equation}
where $x_i$ denotes the transmission data for the $i$th UE with $E\{|x_i|^2\}=1$,
$\mathcal{C} \subseteq \mathcal{L} $ is the set of serving RRHs, $\mathbf{h_{ij}}\in \mathbb{C}^{K\times1} $ denotes the channel vector from RRH $j$ to UE $i$, while $\sigma_i$ denotes the white Gaussian noise which is assumed to be distributed as $\mathcal{CN}(0, \sigma_i^2)$.
Denote $\mathbf{v_{ij}}\in \mathbb{C}^{K\times1} $ as the transmitting beamforming vector from RRH $j$ to UE $i$.
Therefore, the signal-to-interference-plus-noise ratio (SINR) can be expressed by
\begin{equation}\label{en6}
\begin{aligned}
\text{SINR}_i=\frac{|\sum_{j\in \mathcal{C}}\mathbf{v_{ij}}^H \mathbf{h_{ij}}|^2}{\sum_{k\neq i}^N|\sum_{j\in \mathcal{C}}\mathbf{v_{kj}}^H \mathbf{h_{kj}}|^2 +\sigma^2 },    \;\;  i=1, 2,...,N.
\end{aligned}
\end{equation}
Then, the system capacity and the achievable rate for UE $i$ can be given as
\begin{equation}\label{en7}
\begin{aligned}
r_i =B_i \text{log}\left(1+\text{SINR}_i\right)  ,    \;\;  i=1, 2,...,N
\end{aligned}
\end{equation}
where $B_i$ is the wireless channel bandwidth assigning to UE $i$.

The time cost in sending the execution results back to UE $i$ from the RRHs is given by
\begin{equation}\label{en8}
\begin{aligned}
T_i^{Tr}=\frac{D_i}{r_i},  \;\;i=1, 2,...,N
\end{aligned}
\end{equation}
where $D_i$ is the returning data, introduced by the first subsection.
Also, we can assume the power to send this task by RRHs is $p_i$, then the energy consumed by the serving RRHs is
\begin{equation}\label{en9}
\begin{aligned}
E_i^{Tr}=p_i \cdot T_i^{Tr} =\frac{p_i D_i}{r_i},  \;\;i=1, 2,...,N
\end{aligned}
\end{equation}
where $p_i$ can be given as $p_i=\sum_{j\in \mathcal{C}}|\mathbf{v_{ij}}|^2$. Also, we can assume that each RRH $j$ has its own power constraint as follows
\begin{equation}\label{en10}
\begin{aligned}
\sum^{N}_{i=1}|\mathbf{v_{ij}}|^2 \leq P_j, \;\;\;j=1, 2,...,L.
\end{aligned}
\end{equation}

\subsection{Fronthaul Constraints}
The fronthaul link can carry the task results from the mobile clone to the UE through C-RAN. Reference \cite{7037442} uses $\boldsymbol{l} 0 $-norm to model the $j$-th fronthaul capability as
\begin{equation}\label{ene1}
\begin{aligned}
\bar{C}_j=\sum^{N}_{i=1}|\left|\mathbf{v_{ij}}\right|^2|_0,    \;\;\;j=1, 2,...,L
\end{aligned}
\end{equation}
where $|\left|\mathbf{v_{ij}}\right|^2|_0$ denotes the $\boldsymbol{l} 0 $-norm of vector $\left|\mathbf{v_{ij}}\right|^2$, which can be explained as the number of nonzero entries in the vector and also can be mathematically expressed as
\begin{equation}\label{en12}
\begin{aligned}
|\left|\mathbf{v_{ij}}\right|^2|_0=\left\{\begin{matrix}
 &\;\;\;0,  \text{if} \left|\mathbf{v_{ij}}\right|^2 =0 \\& 1,  \text{otherwise}
\end{matrix}\right..
\end{aligned}
\end{equation}
One can see that the number of non-zeros elements of the transmitting beamforming vector $\left|\mathbf{v_{ij}}\right|^2$ also indicates the number of data symbol streams, carried by the fronthaul link from BBU to RRH $j$ for the $i$-th mobile user.
Reference \cite{7037442} also assume that each fronthaul link is only capable of carrying at most $\bar{C}_{j,max}$ signals for UEs as
\begin{equation}\label{ene2}
\begin{aligned}
\bar{C}_j \leq \bar{C}_{j,max},    \;\;\;j=1, 2,...,L.
\end{aligned}
\end{equation}
Reference \cite{6920005} goes a step further and assume that the fronthaul consumption is the accumulated data rates of the users served by RRHs and model the $j$-th fronthaul capability as
\begin{equation}\label{en11}
\begin{aligned}
C_j=\sum^{N}_{i=1}|\left|\mathbf{v_{ij}}\right|^2|_0 \cdot r_i,    \;\;\;j=1, 2,...,L.
\end{aligned}
\end{equation}
In this case, the $j$-th fronthaul constraint can be modeled as the maximum data rates which can be allowed to transmitting through BBU to $j$-th RRH as $C_j \leq C_{j,max}$. Since this constraint is more realistic,
we also use it as the fronthaul constraint in the following derivation of the optimization problem.

\subsection{QoS Requirement}
The QoS can be given as the constraints of the whole time cost for completing the required task and returning the results back to the mobile user. We define the total time spent in executing and transmitting the task results to UE $i$ as
\begin{equation}\label{en15}
\begin{aligned}
T_i=T_i^{Tr}+T_i^C,  \;\;i=1, 2,...,N.
\end{aligned}
\end{equation}
We assume that the task has to be accomplished in time constraints $T_{i,max}$ in order to satisfy the mobile user's requirement, then the QoS can be given as
\begin{equation}\label{en16}
\begin{aligned}
T_i \leq T_{i,max},  \;\;i=1, 2,...,N.
\end{aligned}
\end{equation}

Also, the whole energy cost in executing this task and transmitting the results back to $i$-th UE can be given as
\begin{equation}\label{en14}
\begin{aligned}
E_i= E_i^C+ \eta_i E_i^{Tr},  \;\;i=1, 2,...,N
\end{aligned}
\end{equation}
where $\eta_i \geq 0$ is a weight to trade off between the energy consumptions in the mobile cloud and the C-RAN, and it can be also explained as the inefficiency coefficient of the power amplifier at RRH.

\section{Problem Formulation and separate solutions}
In this section, we provide the energy minimization problem formulation. Our design aims to minimize the energy cost while satisfying the time constraints. First, we formulate the energy minimization for the mobile clone and then we formulate the energy minimization for C-RAN with the fronthaul constraints. Two separate solutions are provided to the energy minimization to the mobile clone and to C-RAN, respectively.

\subsection{Energy Minimization for Mobile Clone}
We assume the time constraint for completing the task in mobile clone as $T_{i,max}^C$, then the energy minimization optimization problem for the mobile clone can be given as
\begin{equation}\label{en17}
\begin{aligned}
\mathcal{P}1: \;\;\; &\underset{f_i^C}{\text{minimize}} \;\;\;   \sum^{N}_{i=1} E_i^C
\\& \text{subject to }:    T_i^C  \leq T_{i,max}^C,
\\& f_i^C \leq  f_{i,max}^C, i=1, 2,...,N.
\end{aligned}
\end{equation}
Assume $f_i^{C^*}$ as the optimum solution for problem $\mathcal{P}1$.
Then, if $f_i^{C^*} \leq f_{i,max}^C$ for $i=1, 2,...,N$, the equality holds for the first constraints. Therefore, the optimal solution can be given by
\begin{equation}\label{en18}
\begin{aligned}
f_i^{C^*}=\frac{F_i}{T_{i,max}^C},\;\;\;\;\;i=1, 2,...,N.
\end{aligned}
\end{equation}
If $f_i^{C^*} > f_{i,max}^C$, we assume there is no solution for the above problem. Thus, the only way to guarantee the QoS is to increase the maximum computation capacity $f_{i,max}^C$ in the cloud.
Therefore, the whole energy cost is given by
\begin{equation}\label{en19}
\begin{aligned}
\begin{cases}
& \sum^{N}_{i=1} \kappa_i^C \frac{F_i^{\nu^L_i}}{(T_{i,max}^C)^{\nu^L_i-1}},\;\;\;\;
\text{if} \;\;\; f_i^{C^*} \leq f_{i,max}^C,  \\
 &  \text{no solution},    \text{ if } \;\;\; f_i^{C^*}  >   f_{i,max}^C,  \;\;i=1, 2,...,N.
\end{cases}
\end{aligned}
\end{equation}

\subsection{Energy Minimization for C-RAN}
We assume the time constraint for transmitting the task results through C-RAN to UE $i$ as $T_{i,max}^{Tr}$. Then, the energy minimization optimization problem for the C-RAN transmission can be given as
\begin{equation}\label{ene3}
\begin{aligned}
\mathcal{P}2: \;\;\;& \underset{\mathbf{v_{ij}}, r_i, \mathcal{C}
}{\text{minimize}}\;\;\;  \sum^{N}_{i=1}  E_i^{Tr}
\\& \text{subject to }: \;\;\; \sum^{N}_{i=1}|\mathbf{v_{ij}}|^2 \leq P_j,
\\&  \sum^{N}_{i=1}|\left|\mathbf{v_{ij}}\right|^2|_0 \cdot r_i  \leq  C_{j,max},
\\&   T_i^{Tr} \leq T_{i,max}^{Tr},  \;\;i=1, 2,...,N, j=1, 2,...,L.
\end{aligned}
\end{equation}
Problem $\mathcal{P}2$ is a non-convex optimization and NP-hard, which is very difficult to solve.
Reference \cite{7130662} has shown that the energy minimization optimization can be probably transformed to the power minimization under some conditions. 
In this subsection, we use some approximations to deal with the energy minimization.

From (\ref{en6}) and (\ref{en7}), one can get the achievable rate for $i$-th UE as
\begin{equation}\label{eene4}
\begin{aligned}
&r_i =B_i \text{log}\left(1+\frac{|\sum_{j\in \mathcal{C}} \mathbf{h_{ij}}^H\mathbf{v_{ij}}|^2}{\sum_{k=1,\; k\neq i}^N | \sum_{j\in \mathcal{C}} \mathbf{h_{ij}}^H\mathbf{v_{kj}}|^2 +\sigma^2}\right), \\&i=1, 2,...,N.
\end{aligned}
\end{equation}
If one ignores the interference term $\sum_{k=1,\; k\neq i}^N | \sum_{j\in \mathcal{C}} \mathbf{h_{ij}}^H\mathbf{v_{kj}}|^2$ and apply Cauchy-Schwarz inequality \cite{Matrix}, one may get
\begin{equation}\label{eene5}
\begin{aligned}
 r_i \leq   B_i \text{log}\left(1+\frac{\sum_{j\in \mathcal{C}} |{\mathbf{h_{ij}}^H}|^2  P_i}{\sigma^2}\right), i=1, 2,...,N.
\end{aligned}
\end{equation}
Then problem $\mathcal{P}2$ may be approximated as \cite{7105959}
\begin{equation}\label{eene3}
\begin{aligned}
\mathcal{P}3: \;\;\;& \underset{\mathbf{v_{ij}}, r_i, \mathcal{C}
}{\text{minimize}}\;\;\;  \sum^{N}_{i=1}  P_i^{Tr}
\\& \text{subject to }: \;\;\; \text{constraints of}  \;(\mathcal{P}2),
\end{aligned}
\end{equation}
where
\begin{equation}\label{ene22}
\begin{aligned}
P^{Tr}_i=\frac{\sum_{j\in \mathcal{C}}|\mathbf{v_{ij}}|^2 D_i}{B_i \text{log}\left(1+\frac{\sum_{j\in \mathcal{C}} |{\mathbf{h_{ij}}^H}|^2  P_i}{\sigma^2}\right)}.
\end{aligned}
\end{equation}
In this case, the equality holds for the last constraint of $\mathcal{P}2$ and then, the minimum transmission data rate can be given by
\begin{equation}\label{en20}
\begin{aligned}
r_i \geq \frac{D_i}{T_{i,max}^{Tr}},  \;\;i=1, 2,...,N.
\end{aligned}
\end{equation}

As the arbitrary phase rotation of the beamforming vectors $\mathbf{v_{ij}}$ does not affect $\mathcal{P}3$, the second constraint of $\mathcal{P}3$ can be rewritten as a second-order cone (SOC) constraint as follows \cite{1561584}
\begin{equation}\label{of22}
\begin{aligned}
&\sqrt{1-\frac{1}{2^{ \frac{D_i }{B_i \cdot T_{i,max}^{Tr}}}}}\sqrt{\sum_{k=1}^N|\sum_{j\in \mathcal{C}} \mathbf{h_{ij}}^H\mathbf{v_{kj}}|^2 +\sigma^2}
\\& \leq \text{Re}\left(|\sum_{j\in \mathcal{C}}\mathbf{h_{ij}}^H\mathbf{v_{ij}}|^2 \right), \;\;  i=1, 2,...,N.
\end{aligned}
\end{equation}

Also, according to \cite{Emmanuel}, the non-convex $\boldsymbol{l} 0 $-norm can be approximated by a convex
reweighted $\boldsymbol{l} 1 $-norm as $|\mathbf{V}|_0=\sum^{N}_{k=1}\rho_{k}|v_k|$, where $v_k$ is the $k$-th element of the vector $\mathbf{V}$ and $\rho_{k}$ is the corresponding weight. Following reference \cite{6920005}, the second last constraint in $\mathcal{P}2$ can be rewritten as follows
\begin{equation}\label{en23}
\begin{aligned}
C_j=\sum^{N}_{i=1}\rho_{ij} \left|\mathbf{v_{ij}}\right|^2 \cdot r_i \leq C_{j,max},    \;\;\;j=1, 2,...,L
\end{aligned}
\end{equation}
where
\begin{equation}\label{en24}
\begin{aligned}
\rho_{ij}=\frac{1}{\left|\mathbf{v_{ij}}\right|^2+\epsilon }
\end{aligned}
\end{equation}
and $\epsilon$ is a small positive factor to ensure stability and can be set as $\epsilon=10^{-10}$ \cite{6920005}.
Then $\mathcal{P}3$ can be transferred to
\begin{equation}\label{ene4}
\begin{aligned}
\mathcal{P}4: \;\;\;& \underset{\mathbf{v_{ij}}, r_i, \mathcal{C}
}{\text{minimize}}\;\;\;  \sum^{N}_{i=1}P^{Tr}_i
\\& \text{subject to }: \;\;\; \sum^{N}_{i=1}|\mathbf{v_{ij}}|^2 \leq P_j,
\\&  \sqrt{1-\frac{1}{2^{ \frac{D_i }{B_i \cdot T_{i,max}^{Tr}}}}}\sqrt{\sum_{k=1}^N|\sum_{j\in \mathcal{C}} \mathbf{h_{ij}}^H\mathbf{v_{kj}}|^2 +\sigma^2}
\\& \leq \text{Re}\left(|\sum_{j\in \mathcal{C}}\mathbf{h_{ij}}^H\mathbf{v_{ij}}|^2 \right),
\\&  C_j=\sum^{N}_{i=1}\rho_{ij} \left|\mathbf{v_{ij}}\right|^2 \cdot r_i \leq C_{j,max},  \\&  \;\;i=1, 2,...,N, j=1, 2,...,L.
\end{aligned}
\end{equation}
Note that by using (\ref{en24}), those beamforming vector $\mathbf{v_{ij}}$ from RRH $j$ to UE $i$ with lower values will have higher weights $\rho_{ij}$, and will be further forced to reduce and finally be encouraged to become zero. In this process. RRH cluster could be formed to serve its corresponding UE \cite{6920005}. This is how we obtain $\mathcal{C}$ in this paper.

Note also that $\mathcal{P}4$ without the fronthaul constraint is an SOC problem, which can be solved by the interior-point method \cite{Stephen}, while $\mathcal{P}4$ including the fronthaul constraint can be addressed by the iterative solution, as shown in \cite{6920005}. Therefore we can give the iterative Algorithm 1 to deal with $\mathcal{P}4$, where
\begin{equation}\label{enee5}
\begin{aligned}
P^{Tr}=\sum^{N}_{i=1}P_i^{Tr}.
\end{aligned}
\end{equation}
\begin{table}[htbp]
 \begin{center}
 \begin{tabular}{ll}
  \toprule
  \textbf{Algorithm 1} \;\;\;  Proposed iterative algorithm for $\mathcal{P}4$   \\
  \midrule
  \textbf{Initialize}: \;\;\; $m=1$, $\rho_{ij}^{(0)}=0$, $r_i^{(0)}=1$, $i=1, 2,...,N$, \\ $j=1, 2,...,L$;\\
  \textbf{Repeat}:  \\
  $1$: \;\;\;  Solve the second-order cone programming (SOCP)\\ optimization $\mathcal{P}4$ using interior-point method, \\obtaining the
  optimal beamforming vector $\mathbf{v_{ij}}^{(m)}$;  \\
  $2$: \;\;\;  Update $r_i^{(m+1)} = r_i^{(m)}$ according to (\ref{eene4}); \\
  $3$: \;\;\;  Update $\rho_{ij}^{(m+1)} = \rho_{ij}^{(m)}$ according to (\ref{en24}); \\
  $4$: \;\;\;  Update $P^{{Tr}^{(m+1)}} = P^{{Tr}^{(m)}}$ according to (\ref{ene22}) and (\ref{enee5}); \\
  $5$: \;\;\;  $m=m+1$; \\
  \textbf{Until} $|P^{{Tr}^{(m+1)}}-P^{{Tr}^{(m)}}|<\varepsilon$, or maximum number \\  of iterations is reached. \\
  \textbf{Return}: RRH cluster $\mathcal{C}$, beamforming vector $\mathbf{v_{ij}}$ and \\date rate $r_i$,
  for $i=1, 2,...,N$,  $j=1, 2,...,L$. \\
  \bottomrule
 \end{tabular}
 \end{center}
\end{table}

One can see that the computational complexity of Algorithm 1 mostly come from the Step 1, i,e., SOCP optimization, which can be solved by interior-point method.
Suppose Algorithm 1 needs $M$ total number of iterations to converge or the maximum number of iterations is set to $M$, then the computational complexity can be approximately given as $O(M\cdot(KNL)^{3.5} )$ \cite{YYe}.
\section{Joint Optimization Solution}
In this section, we will solve the energy minimization optimization and resource allocation jointly between the mobile cloud and mobile network. The objective is to minimize the total energy consumption in mobile cloud for executing the task and in C-RAN for transmitting the processing results back to the mobile user. We assume that the task has to be completed in the total time constraint (QoS) of the given task, including the executing time plus the transmitting time.
Therefore, the joint energy optimization problem can be given as
\begin{equation}\label{en25}
\begin{aligned}
\mathcal{P}5: \;\;\;& \underset{f_i^C, r_i, \mathbf{v_{ij}}, \mathcal{C}
}{\text{minimize}}\;\;\;  \sum^{N}_{i=1} E_i
\\& \text{subject to }:
\\& \sum^{N}_{i=1}|\mathbf{v_{ij}}|^2 \leq P_j,
\\& f_i^C \leq  f^C_{i,max},
\\& T_i^C+T_i^{Tr} \leq T_{i,max },
 \\& \sum^{N}_{i=1}|\left|\mathbf{v_{ij}}\right|^2|_0 \cdot r_i  \leq  C_{j,max},    \\&  \;\;i=1, 2,...,N, j=1, 2,...,L
\end{aligned}
\end{equation}
where $r_i$ is given by (\ref{eene4}), $E_i=E_i^C+ \eta_i E_i^{Tr}$, and other constraints in $\mathcal{P}5$ have been introduced in the last sections.
The above $\mathcal{P}5$ is non-convex problem and difficult to solve. In the next subsections, we will provide the iterative algorithms based on WMMSE solution to deal with it.

\subsection{Problem Transformation}
Following the same process before, $\mathcal{P}5$ can be approximated as
\begin{equation}\label{een25}
\begin{aligned}
& \underset{f_i^C, r_i, \mathbf{v_{ij}}, \mathcal{C}
}{\text{minimize}}\;\;\;  \sum^{N}_{i=1} \kappa_i^C (f_i^C)^{\nu^C_i-1} F_i \\&+\eta_i \frac{\sum_{j\in \mathcal{C}}|\mathbf{v_{ij}}|^2 D_i}{B_i \text{log}\left(1+\frac{\sum_{j\in \mathcal{C}} |{\mathbf{h_{ij}}^H}|^2  P_i}{\sigma^2}\right)}
\\& \text{subject to}: \text{constraints of}  \;(\mathcal{P}5).
\end{aligned}
\end{equation}
Then, the equality of the time constraint holds for $\mathcal{P}5$ in relaxation. Therefore, by using (\ref{en2}) and (\ref{en8}), time constraint may be relaxed as
\begin{equation}\label{en26}
\begin{aligned}
T_{i,max} &= T_i^{Tr}+T_i^C
\\&=\frac{D_i}{r_i} +\frac{F_i}{f_i^C},  \;\;i=1, 2,...,N.
\end{aligned}
\end{equation}
Then, $f_i^C$ can be written as
\begin{equation}\label{en27}
\begin{aligned}
f_i^C=\frac{F_i}{T_{i,max}-\frac{D_i}{r_i}},  \;\;i=1, 2,...,N.
\end{aligned}
\end{equation}
Given that $T_{i,max} > 0$, $f_i^C>0$ and $f_i^C \leq  f^C_{i,max}$, one can get the minimum achievable rate as
\begin{equation}\label{en28}
\begin{aligned}
r_i \geq R_{i,min},
\end{aligned}
\end{equation}
where
\begin{equation}\label{en29}
\begin{aligned}
R_{i,min}=\frac{D_i}{T_{i,max}-\frac{F_i}{f^C_{i,max}}},  \;\;i=1, 2,...,N.
\end{aligned}
\end{equation}
We denote $\mathbf{v_j}=[\mathbf{v_{1j}}, \mathbf{v_{2j}},..., \mathbf{v_{Nj}}]^H $,
$\mathbf{h_j}=[\mathbf{h_{1j}}, \mathbf{h_{2j}},..., \mathbf{h_{Nj}}]^H $,
$\mathbf{v_i}=[\mathbf{v_{i1}}, \mathbf{v_{i2}},..., \mathbf{v_{i\mathcal{C}}}]^H $ and
$\mathbf{h_i}=[\mathbf{h_{i1}}, \mathbf{h_{i2}},..., \mathbf{h_{i\mathcal{C}}}]^H $ for notation simplification. By using (\ref{en27}), (\ref{en28}) and (\ref{en29}), $\mathcal{P}5$ can be rewritten as
\begin{equation}\label{en30}
\begin{aligned}
\mathcal{P}6: \;\;\;& \underset{ r_i, \mathbf{v_{ij}}, \mathcal{C}
}{\text{minimize}}\;\;\; \sum^{N}_{i=1} \sum^{N}_{i=1} \gamma_i(r_i)  + \beta_i (\mathbf{\mathbf{v_i}})
\\& \text{subject to }:
 \\& \sum^{N}_{i=1}|\mathbf{v_{ij}}|^2 \leq P_j,
\\&  r_i \geq R_{i,min},
\\&  C_j=\sum^{N}_{i=1}\rho_{ij} \left|\mathbf{v_{ij}}\right|^2 \cdot r_i \leq C_{j,max},    \\&  \;\;i=1, 2,...,N, j=1, 2,...,L
\end{aligned}
\end{equation}
where
\begin{equation}\label{en32}
\begin{aligned}
\gamma_i(r_i)  = \kappa_i^C \left(\frac{F_i}{T_{i,max}-\frac{D_i}{r_i}}\right)^{\nu^C_i-1} F_i
\end{aligned}
\end{equation}
and
\begin{equation}\label{en34}
\begin{aligned}
\beta_i (\mathbf{\mathbf{v_i}}) =\eta_i \frac{\mathbf{v_i}^H \mathbf{v_i} D_i}{ B_i \text{log}\left(1+\frac{\sum_{j\in \mathcal{C}} |{\mathbf{h_{ij}}^H}|^2  P_i}{\sigma^2}\right)  }.
\end{aligned}
\end{equation}
Note that $f_i^C$ does no longer exist in $\mathcal{P}6$, which can be solved by using WMMSE-based iterative solution shown in the next subsection.

\subsection{WMMSE-based Solution}
One can see that the objective of $\mathcal{P}6$ is a decreasing function of the mobile user's data rate $r_i$. Also, one can recall the well-known relation between MSE covariance matrix and the rate $r_i$ as follows
\begin{equation}\label{ene36}
\begin{aligned}
r_i = \text{log}\left({(e_i)}^{-1}\right),  \;\;i=1, 2,...,N.
\end{aligned}
\end{equation}
Then, the sum rate maximization problem can be transformed to the weighted sum MSE minimization optimization solved by WMMSE method \cite{5756489,4712693}.
Thus, one can reformulate $\mathcal{P}8$ as an equivalent WMMSE problem and use the block coordinate descent approach to deal with it.

Assume the receiving beamforming vector in mobile user $i$ as $\mathbf{u_i} \subseteq \mathbb{C}^{1\times1}$, as there is only one antenna in the UE. Thus, the corresponding MSE at UE $i$ can be given as
\begin{equation}\label{en37}
\begin{aligned}
e_i=& E \left[ (\mathbf{u_i}y_i-x_i)  (\mathbf{u_i}y_i-x_i)^H \right]
\\&=\sum^{N}_{i=1} \mathbf{u_i}^H( \mathbf{h_i}^H\mathbf{v_i}\mathbf{v_i}^H \mathbf{h_i} +\sigma_i^2)\mathbf{u_i}
-2\; \text{Re}\left[ \mathbf{u_i}^H  \mathbf{h_i}^H\mathbf{v_i} \right]+1, \\& \;\;i=1, 2,...,N.
\end{aligned}
\end{equation}
Then, $\mathcal{P}6$ can be transformed to
\begin{equation}\label{w44}
\begin{aligned}
\mathcal{P}7: \;\;\; &\underset{\phi_i, \mathbf{v_{ij}}, \mathbf{u_i}, \mathcal{C}}{\text{minimize}} \;\;\;   \sum^{N}_{i=1}   \phi_i e_i+ \tau_i (\omega_i (\phi_i) )- \phi_i (\omega_i (\phi_i)) + \\& \beta_i (\mathbf{v_i})
\\& \text{subject to}:  \text{constraints of}  \;(\mathcal{P}6)
\end{aligned}
\end{equation}
where
\begin{equation}\label{en39}
\begin{aligned}
\tau_i (e_i) = \gamma_i(-B_i \cdot \text{log}(e_i) ),
\end{aligned}
\end{equation}
and $\omega_i(\cdot)$ is the inverse mapping of the gradient map $\frac{\partial  \tau_i(e_i)}{\partial e_i}$. One can see that $\tau_i(e_i)$ is a strictly concave function in $\mathcal{P}7$, as $\gamma_i (r_i)$ is the decreasing utility function of the data rate $r_i$.
One can see that $\mathcal{P}7$ is convex with respect to each of the individual variables $\phi_i$, $\mathbf{v_{ij}}$ and $\mathbf{u_i}$.
Therefore, one can use the block coordinate descent method to solve it \cite{7105959}, \cite{6920005}, \cite{5756489}, \cite{4712693}.
The process to solve $\mathcal{P}7$ is as follows:

\textbf{Step 1:}
By fixing all the transmit beamforming vector $\mathbf{v_i}$, the optimal receive beamforming vector can be give by
the well-known minimum mean square error (MMSE) receiver as
\begin{equation}\label{en38}
\begin{aligned}
&\mathbf{u_i}= \left(\mathbf{h_i}^H\mathbf{v_i} \right)
\cdot \left(  \sum_{k=1}^N  \mathbf{h_i}^H\mathbf{v_k} \mathbf{v_k}^H\mathbf{h_i}  + \sigma_i^2 \right)^{-1},\\&  \;\;i=1, 2,...,N.
\end{aligned}
\end{equation}

\textbf{Step 2:}
By fixing the transmit beamforming vector $\mathbf{v_i}$ and the MMSE receiver $\mathbf{u_i}$, the corresponding optimal MSE weight $\phi_i$ can be given by
\begin{equation}\label{en42}
\begin{aligned}
&\phi_i=\frac{\partial \tau (e_i)}{\partial e_i} \\& = \frac{D_i \kappa_i^C  (\nu_i^C-1) \log (2) \left(\frac{B_i F_i \log (e_i )}{B_i T_{i,max}\log (e_i )+D_i \log (2)}\right)^{\nu_i^C}}{B_i e_i  \log ^2(e_i)}, \\& \;\;i=1, 2,...,N.
\end{aligned}
\end{equation}

\textbf{Step 3:}
By fixing the optimal MSE weight $\phi_i$ and MMSE receiver $\mathbf{u_i}$, the optimal transmit beamforming vector $\mathbf{v_i}$ can be calculated by solving the following quadratically constrained quadratic programming (QCQP), which can also be transformed to
SOCP as
\begin{equation}\label{ene5}
\begin{aligned}
&\underset{ r_i,  \mathbf{v_{ij}}, \mathcal{C}
}{\text{minimize}}\;\;\; \sum^{N}_{i=1} \phi_i \cdot e_i  + \beta_i (\mathbf{v_i})
\\& \text{subject to}: \text{constraints of}  \;(\mathcal{P}6).
\end{aligned}
\end{equation}

Thus, we can deal with the overall optimization problem with WMMSE-based iterative method as in Algorithm 2,
where $\varepsilon$ is a small constant to guarantee convergence and
\begin{equation}\label{e5}
\begin{aligned}
E=\sum^{N}_{i=1}E_i.
\end{aligned}
\end{equation}
\begin{table}[htbp]
 \begin{center}
 \begin{tabular}{ll}
  \toprule
  \textbf{Algorithm 2} \;\;\;  Proposed iterative algorithm for joint \\ optimization problem   \\
  \midrule
  \textbf{Initialize}: \;\;\; $n=1$, $\rho_{ij}^{(0)}=1$, $r_i^{(0)}=1$, $\mathbf{v_{ij}}^{(0)}$, \\$i=1, 2,...,N$, $j=1, 2,...,L$. \\
  \textbf{Repeat}:  \\
  $1$: \;\;\;  Obtain the receive beamforming vector $\mathbf{u_i}^{(n)}$ \\ according to (\ref{en38}) by fixing $\mathbf{v_{ij}}^{(n-1)}$; \\
  $2$: \;\;\;  Obtain the MSE weight $\phi_i$ according to (\ref{en42}) \\by fixing $\mathbf{v_{ij}}^{(n-1)}$ and $\mathbf{u_i}^{(n)}$;\\
  $3$: \;\;\;  Obtain the transmit beamforming vector $\mathbf{v_{ij}}^{(n)}$ \\according to SOCP (\ref{ene5}) by fixing $\phi_i^{(n)}$, $\mathbf{u_i}^{(n)}$; \\
  $4$: \;\;\;  Update $r_i^{(n+1)} = r_i^{(n)}$ according to (\ref{eene4}); \\
  $5$: \;\;\;  Update $\rho_{ij}^{(n+1)} = \rho_{ij}^{(n)}$ according to (\ref{en24}); \\
  $6$: \;\;\;  Update $E^{(n+1)} = E^{(n)}$ according to (\ref{e5}); \\
  $7$: \;\;\;  $n=n+1$; \\
  \textbf{Until} $|E^{(n+1)}-E^{(n)}|<\varepsilon$, or maximum number \\  of iterations is reached. \\
  \textbf{Return}: RRH cluster $\mathcal{C}$, beamforming vector $\mathbf{v_{ij}}$, \\date rate $r_i$, and computational capacity $f_i$,
  \\for $i=1, 2,...,N$, $j=1, 2,...,L$. \\
  \bottomrule
 \end{tabular}
 \end{center}
\end{table}

One can see that the computational complexity of Algorithm 2 mostly come from the Step 3, i,e., SOCP optimization, which can be solved by interior-point method.
Similar to Algorithm 1, suppose Algorithm 2 needs $M$ total number of iterations to converge or the maximum number of iterations is set to $M$, then the computational complexity can be approximately given as $O(M\cdot(KNL)^{3.5} )$ \cite{YYe}.

\section{Simulation Results}
\begin{figure}[htbp]
\centering
\includegraphics[width=2.3in]{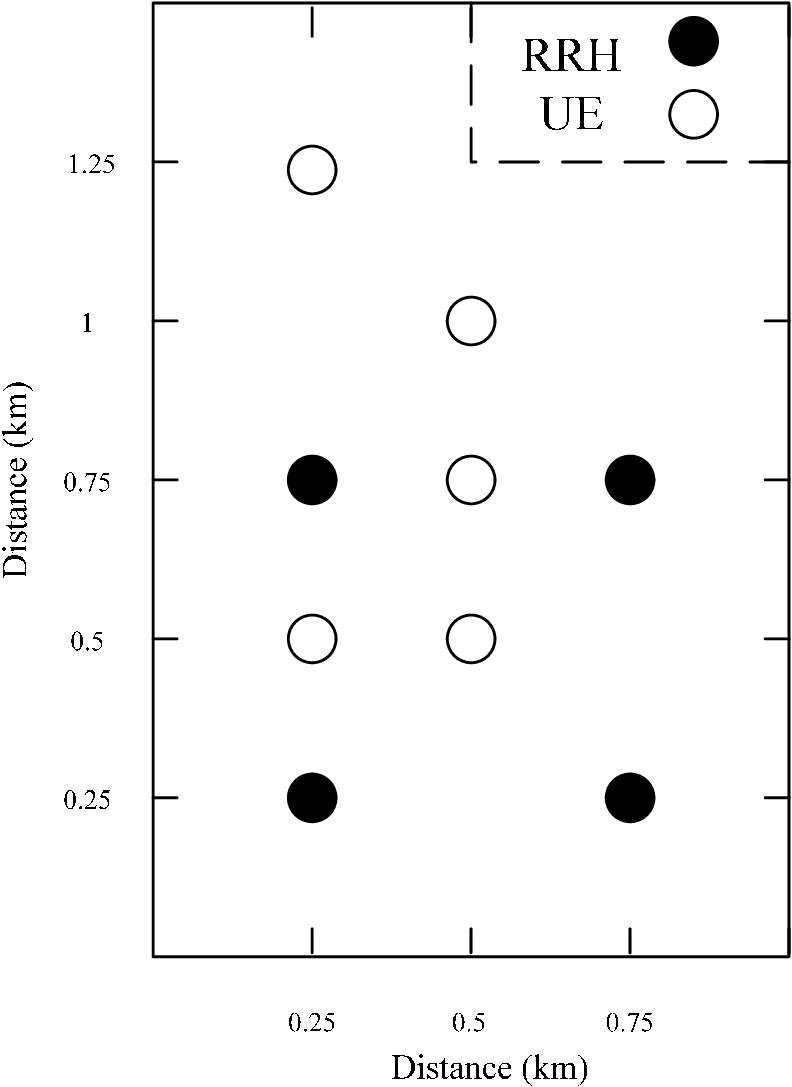}
\caption{C-RAN network with $L=4$ RRHs and $N=5$ UEs.} \label{figf2}
\end{figure}

In this section, simulation results are provided to show the effectiveness of the proposed joint energy minimization optimization.
Matlab with CVX tool \cite{cvx} has been used in the simulation.
The simulation parameters are summarized in Table. \ref{tab1} and the simulation
environment is shown in Fig. \ref{figf2}, in which
we consider the C-RAN network with $L=4$ RRHs, each equipped with $K=2$ antennas. Also, we assume there are $N=5$ mobile users, each of which has only one antenna. We assume there are five mobile clones co-located with the BBUs, and each mobile clone has the same software stack as its corresponding mobile users and can execute the task for the mobile user.
\begin{table}
 \caption{\label{tab1}Simulation Parameters.
 }
\begin{center}
{\begin{tabular}{|c|c|c|c|c|}
\hline
Parameter & Description & Value   \\
\hline
\hline
$L$ & Number of RRHs & 4  \\
\hline
$K$ & Number of antennas of RRH  & 2   \\
\hline
$N$ & Number of UEs  & 5  \\
\hline
$P_j, j\in \mathcal{C}$ & Power constraint for RRH   & 1 W   \\
\hline
$f^C_{i,max}, i\in \mathcal{N}$ &Computation capacity constraint  & 1 M \\
\hline
$\eta_i, i\in \mathcal{N}$ & Trade off factor & 10  \\
\hline
$B_i, i\in \mathcal{N}$ & Bandwidth  & 10 MHz  \\
\hline
$C_{j,max}, j\in \mathcal{C}$ & Fronthaul capacity  & 10 Mbps   \\
\hline
$\nu^C_i, i\in \mathcal{N}$ & Cloud computation parameter & 3   \\
\hline
\end{tabular}}
\end{center}
\end{table}
Moreover, we assume the maximum transmit power for each RRH is $1$ W, while the maximum computation capacity for each mobile clone is 1 M CPU cycles per second.
Similar to \cite{6855687},
we model the path and penetration loss as
\begin{equation}\label{ene6}
\begin{aligned}
p (d)=127+25\text{log}10(d)
\end{aligned}
\end{equation}
where $d \; (km)$ is the propagation distance. Also, we model the small scale fading as independent circularly symmetric Gaussian process distributed as $\mathcal{CN}(0, 1)$, whereas the noise power spectral density is assumed to be $-100\; \text{dBm/Hz}$. We assume the energy tradeoff factor between the mobile clone and C-RAN as $\eta_i=10$, the parameter for the cloud energy model $\nu^C_i=3$ and $\epsilon=10^{-10}$. Also, we assume the wireless channel bandwidth as $10$ MHz and the fronthaul capacity constraint as $10$ Mbps.

\begin{figure}[htbp]
\centering
\includegraphics[width=3.7in]{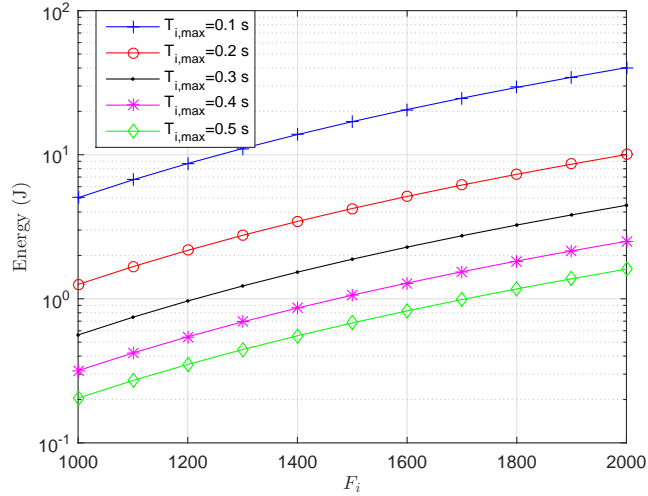}
\caption{Total energy consumption vs. CPU cycles under different $T_{i,max}$ with $D_i=1000$.} \label{fi}
\end{figure}

In Fig. \ref{fi}, we show the energy consumption for the whole system including mobile clone and C-RAN for different QoS requirement and different CPU cycles of the task. Transmission data $D_i=1000$ bits is set in this figure. One can see that with the increase of the CPU cycles of the task $F_i$, the energy cost rise correspondingly. Also, with the increase of the time constraint, the total energy decrease, as the mobile clone and the C-RAN can have more time to complete the task and return the result to the mobile user.

\begin{figure}[htbp]
\centering
\includegraphics[width=3.7in]{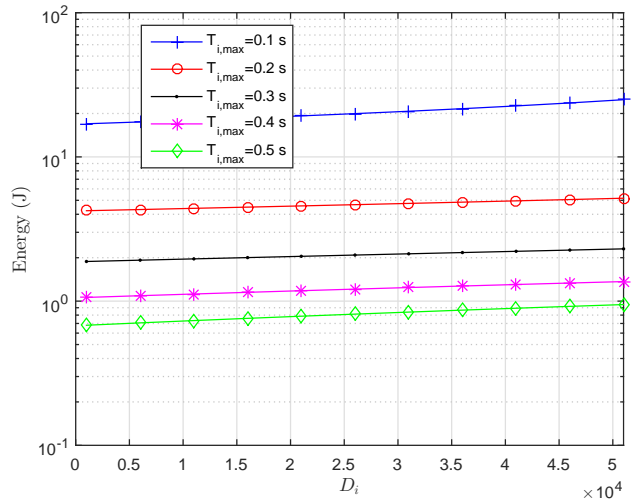}
\caption{Total energy consumption vs. data size under different $T_{i,max}$ with $F_i=1500$.} \label{di}
\end{figure}

In Fig. \ref{di}, we show the total energy consumption for different QoS requirement and different data size of the transmission. $F_i=1500$ CPU cycles is set in this figure. One can see that with the increase of the result data size $D_i$ of the task, the energy cost increase correspondingly, but not as fast as Fig. \ref{fi}. This is due to the tradeoff factors we set.
Similarly to Fig. \ref{fi}, with the increase of the time constraint, the total energy cost decrease. This can be also explained that with the increase of the QoS level, more energy is correspondingly required.

\begin{figure}[htbp]
\centering
\includegraphics[width=3.7in]{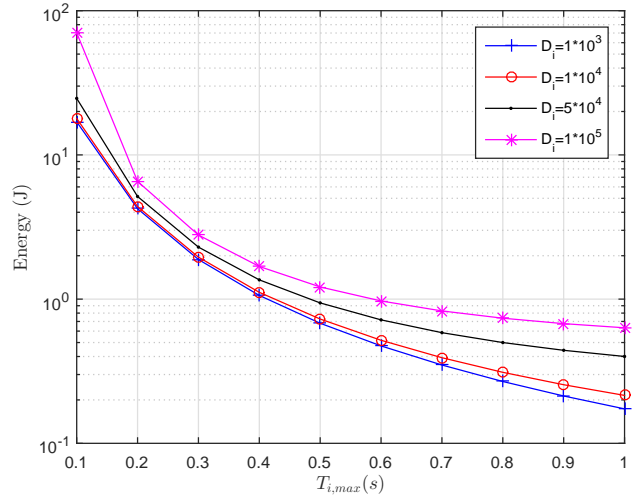}
\caption{Total energy consumption vs. time constraint under different data size $D_i$ with $F_i=1500$.} \label{fidi}
\end{figure}

In Fig. \ref{fidi}, the relations between the total energy consumption and different QoS or time constraints are examined under different $D_i$ with total CPU cycles $F_i=1500$. One can see that with the increase of the time constraints, the energy consumption decreases, as expected. Also, with the increase of the data size, the energy increases, but the gap between them is small, due to the tradeoff factor we set.

\begin{figure}[htbp]
\centering
\includegraphics[width=3.7in]{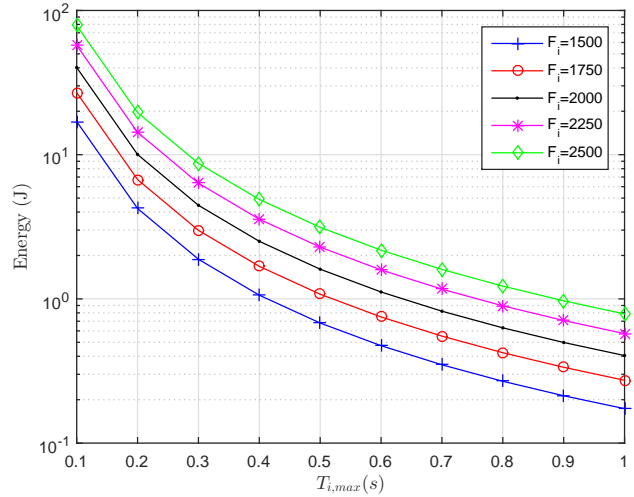}
\caption{Total energy consumption vs. time constraint under different CPU cycles $F_i$ with $D_i=1000$.} \label{tifi}
\end{figure}

Similar to Fig. \ref{fidi}, Fig. \ref{tifi} shows that the whole energy consumption of mobile cloud and C-RAN decreases either with the increase of the time constraints or with the decrease of the CPU cycles required by each task.

\begin{figure}[htbp]
\centering
\includegraphics[width=3.7in]{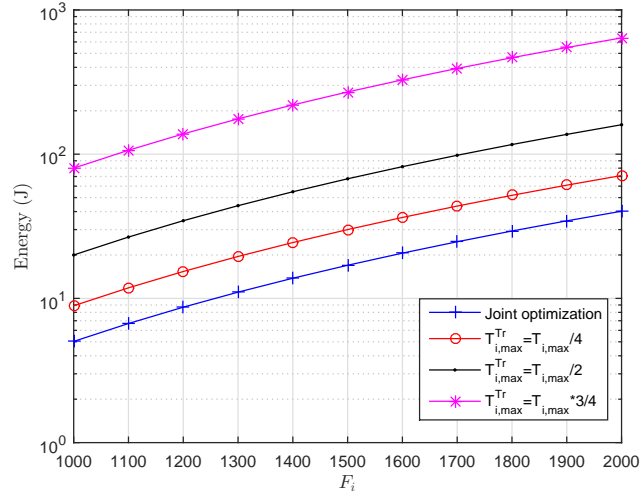}
\caption{Total energy consumption vs. CPU cycles under different $T_{i,max}^{Tr}$ with $D_i=1000$.} \label{comparefi}
\end{figure}

\begin{figure}[htbp]
\centering
\includegraphics[width=3.7in]{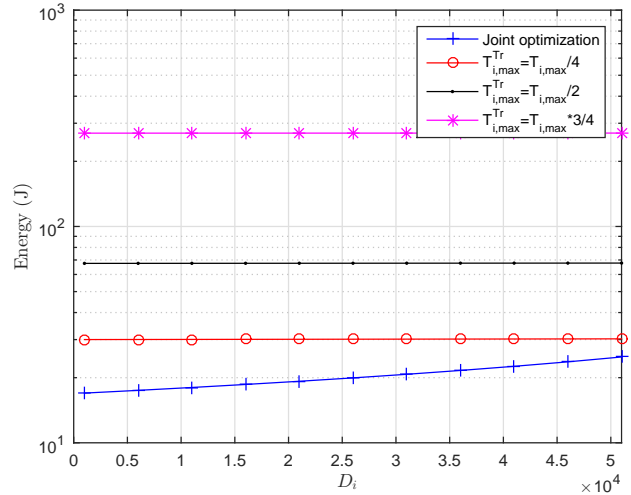}
\caption{Total energy consumption vs. data size under different $T_{i,max}^{Tr}$ with $F_i=1500$.} \label{comparedi}
\end{figure}

In Fig. \ref{comparefi} and Fig. \ref{comparedi}, we compare the proposed joint energy minimization optimization with the separate energy minimization solutions, which has been used in some works such as \cite{6855687}, etc. For the separate energy minimization, we set two time constraints as $T_i^{Tr} \leq T_{i,max}^{Tr}$ and $T_i^{C} \leq T_{i,max}^C$, where $T_{i,max}^{Tr}+T_{i,max}^C=T_{i,max}$. $T_{i,max}=0.1 $s is set in both
Fig. \ref{comparefi} and Fig. \ref{comparedi} while $D_i=1000$ and $F_i=1500$ are set in Fig. \ref{comparefi} and Fig. \ref{comparedi}, respectively. One can see that the joint energy minimization achieves the best performance, followed by the second best solution when setting $T_{i,max}^{Tr}=T_{i,max}^{Tr}/4$ in both Fig. \ref{comparefi} and Fig. \ref{comparedi}. The performance of $T_{i,max}^{Tr}=T_{i,max}^{Tr}*3/4$ can be shown as the worst solution among the test ones in both figures. Therefore, the simulation results show that the proposed joint energy minimization outperforms the separate solutions in all the cases.

\begin{figure}[htbp]
\centering
\includegraphics[width=3.7in]{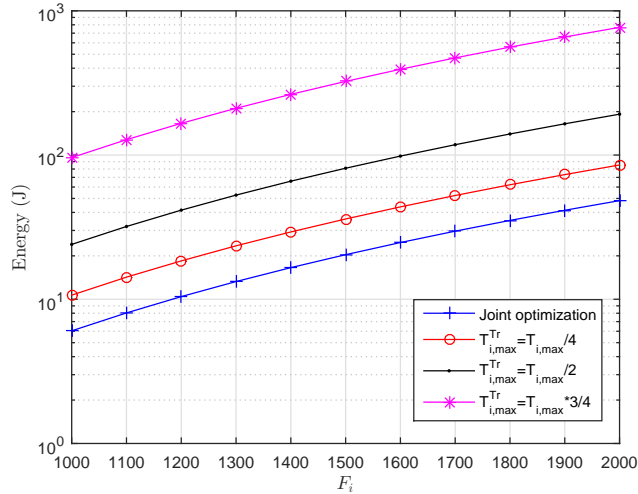}
\caption{Total energy consumption vs. CPU cycles for six mobile users.} \label{addoneuser}
\end{figure}

In Fig. \ref{addoneuser}, we assume that one additional user has been added in C-RAN system in Fig. \ref{figf2} and other parameters are set the same as in Fig. \ref{comparefi}. One can see that our proposed optimization method has nearly the same performance gain as in Fig. \ref{comparefi}. As expected, more power is used for all the solutions in Fig. \ref{addoneuser} than Fig. \ref{comparefi}. Also, we have checked our our solution for different number of antennas and similar performance gain can be achieved. However we do not show those figures due to limited space.

\section{Conclusion}
A novel C-RAN architecture with the mobile clones involved is proposed in this paper by taking full advantages of the two cloud-based techniques. In particular, we assume there is one task needed to be executed in the mobile clone for each UE and we model this task with two features, i.e, the total number of the CPU cycles required to complete this task and the total data size required to transmit the result back to the UEs through C-RAN. We jointly minimize the whole energy cost in mobile cloud and mobile network by modeling this problem into the optimization problem when taking QoS, i.e., the time constraint into consideration. Also, we have considered the fronthaul constraints in C-RAN in order to get the RRH clusters. Numerical results are presented to show that the proposed energy minimization and resource allocation solution can improve the system performance and save energy.

Future work will be focused on the whole data transmission process including the uplink (i.e., the UE sending user data to RRH) and downlink transmission (i.e., the RRH sending result data back to RRH). Also, we aim to model the fronthaul transmission in C-RAN, including transmission time model and energy consumption model in fronthaul.


%

\ifCLASSOPTIONcompsoc
  \section*{Acknowledgments}
\else
  \section*{Acknowledgment}
\fi

This work was supported by UK EPSRC NIRVANA project under the grant No. EP/L026031/1 and EU Horizon 2020 iCIRRUS project under the grant No. GA-644526.

\ifCLASSOPTIONcaptionsoff
  \newpage
\fi



%

\bibliographystyle{ieeetran}
\bibliography{bare_jrnl}


%









\end{document}